# Behaviour Prediction of Closed-loop HTS coils in Non-Uniform AC fields

Zhuoyan Zhong[1], Wei Wu[1, *], Xueliang Wang[1], Xiaofen Li[1], Jie Sheng[1], Zhiyong Hong[1] and Zhijian Jin[1]

[1] Department of Electrical Engineering, Shanghai Jiao Tong University, Shanghai, 200240, People's Republic of China

E-mail: wei.wu@sjtu.edu.cn

Field decay rate is the key characteristic of the superconducting magnets based on closed-loop coils. However, in Maglev trains or rotating machines, closed-loop magnets work in external AC fields and will exhibit an evidently accelerated field decay resulting from dynamic resistances, which are usually much larger than joint resistance. Nevertheless, there has not been a numerical model capable of systematically studying this behaviour, which is the main topic of this work. The field decay curves of a closed-loop high-temperature-superconducting (HTS) coil in various AC fields are simulated based on *H*-formulation. A non-uniform external field generated by armature coils is considered. Reasonable consistence is found between experimental and simulation results. In our numerical model, the impact of current relaxation, which is a historical challenge, is analysed and subsequently eliminated with acceptable precision. Our simulation results suggest that most proportion of the field decay rate is from the innermost and outermost turns. Based on this observation, a magnetic shielding pattern is designed to reduce the field decay rate efficiently. This work has provided magnet designers an effective method to predict the field decay rate of closed-loop HTS coils in external AC fields, and explore various shielding designs.

## 1. Introduction

Various high temperature superconducting (HTS) magnet applications may require closed-loop coils to work in persistent current mode (PCM) to reduce heat losses and generate stable magnetic field [1-3]. Among these applications, field decay performance is the key factor for their operation. In a stationary operation, the field decay rate is determined by the joint resistance and index loss (if with a large working current). However, during the dynamic operation, e.g. rotating machines, Maglev trains and other potential applications [1, 2], the closed-loop coils are exposed to external AC magnetic field. Therefore, dynamic resistance, which arises from the interactions between the transport DC current and moving fluxons [4], is likely to be generated in the closed-loop. In these cases, the field decay of the closed-loop coil is accelerated and exhibits a significant deviation from its original design.

Researchers have studied the experimental field decay of closed-loop magnets under AC field [5-7], as well as the angular dependence of the external field [8]. Although the experimental results suggested that external AC field would accelerate the field decay much more significant than joint resistance [5], this behaviour is still unpredictable.

Dynamic resistance for single or stacked tapes, as a key feature of HTS flux pump devices, have been studied by both analytical expressions [4, 9-13] and numerical models [14, 15], which have been proven to be in good agreement with experimental results. Zhang *et al* has tried to calculate AC loss of HTS coils including dynamic loss [16]. However, dynamic resistance for coils, as the acceleration mechanism of the field decay of closed-loop magnets, has not been systematically studied by numerical method and experiment.

This paper presents a modelling technique for simulating the field decay of a closed-loop GdBCO magnet under AC field, and our experimental verification. The *H*-formulation and *E-J* power law assumption are utilised. Non-uniform external field generated by the armature coils is considered. The challenges are: (1) for coil cases, it is found that the historical challenge involving the eddy current model, current relaxation [17-20], is very influential to the results and must be properly considered; and (2) the induced current circulating in a closed-loop coil, which is temporarily eliminated in this work.

## 2. Numerical Simulation

### 2.1 Geometry and governing equations

A double-pancake (DP) HTS coil and two copper coils were established in a 2D axisymmetric model based on the *H*-formulation and *E-J* power law in COMSOL™ [18]. The copper coils served as armature coils to generate AC fields. The general model geometry is shown in figure 1(a). The resistivity ($\rho$) was set to $8 \times 10^{-9}$ $\Omega \cdot m$ for copper coil regions and 100 $\Omega \cdot m$ for air region.

The anisotropy of our experimental GdBCO tape at 77 K was calculated by equation (1) introduced by [21]. The explanation of P1($B$), P2($B$), and G($\theta$) is illustrated in figure 2.

$$J_c(B, \theta) = J_{c0} * \{P1(B) + [P2(B) - P1(B)] * G(\theta)\} \quad (1)$$

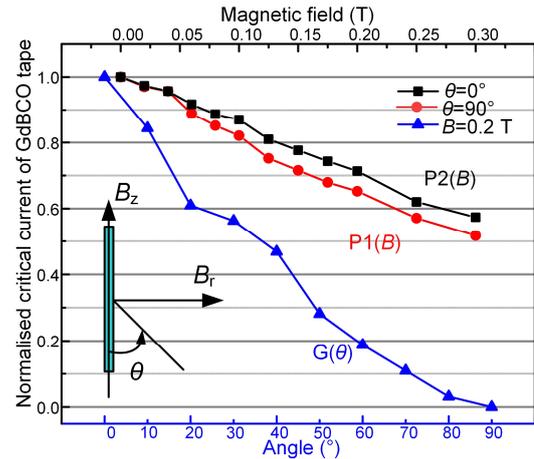

**Figure 2.** Normalised experimental $I_c$ of the GdBCO tape under different perpendicular fields (P1($B$)), parallel fields (P2($B$)), and angles of the 0.2 T field (G($\theta$)).

The homogenisation technique [22] is used to improve the computation speed. The thickness of the superconducting layer is artificially expanded to the thickness of the insulated tape in our experiment in section 4. The measured self-field critical current of the single tape at 77 K, $I_{c0}$, is 213 A. The *n*-value is set as 25. The self-field critical current density, $J_{c0}$, in equation (1), is then modified accordingly to keep $I_{c0}$ unchanged.

### 2.2 Simulation by direct field cooling and decay in an AC field

To fully simulate the excitation and decaying procedures of closed-loop coils, no constraint should be applied to the current of HTS coil. Because the transport current in the closed-loop coil, $I_{DC}$, is the unknown variable. The transport current in HTS coil is excited by the field-cooling method [3] with one of the copper coils. A 50 Hz external field is generated by two copper coils.

A case study is conducted for a single-turn HTS coil. The detailed coil system is shown in section 4; however, the HTS coil is a single-turn coil with an inner radius of 5 cm. As is shown in figure 3, the field-cooling excitation and current decay curves under AC field can be modelled by pure 2D FEM. The field decay curves can be obtained by simply multiplying the coil constant. It is clear that a greater decay occurs after an AC external field is applied at 5.4 s.

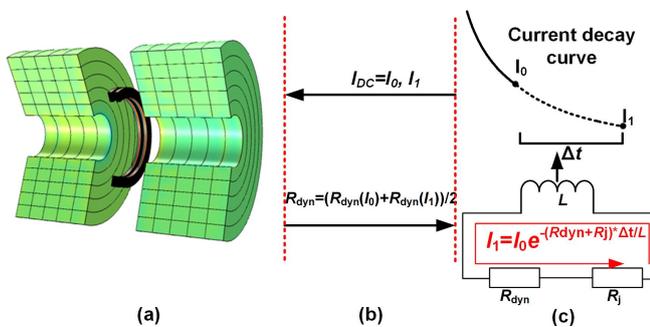

(a)                      (b)                      (c)

**Figure 1.** (a) Geometry of the HTS coil and two copper coils in the *H*-formulation model. (b) Coupling of FEM and circuit model. (c) Current decay curve calculation method in the circuit model.

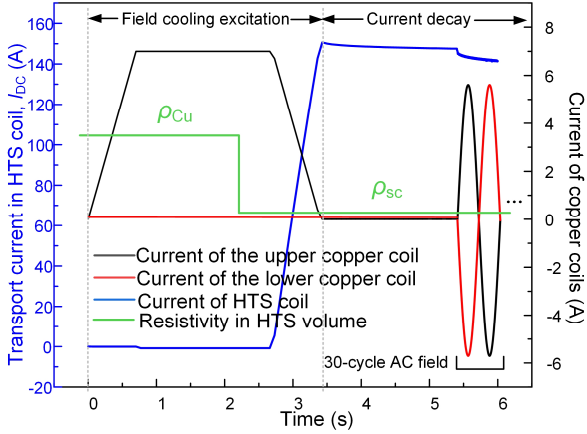

**Figure 3.** Simulated results of the field-cooling operation and current decay of a single-turn closed-loop coil under a 50 Hz field.

However, this simulation is not available for multi-turn coils (but available for ring structure coils), because the current constraint cannot be applied and no effective constraint exists to ensure that the current is identical among turns. Thus, another method should be adopted to calculate the current decay curves of multi-turn coils.

## 2.3 Simulation by the discrete calculation of dynamic resistance

In figure 1(c), the closed-loop coil is regarded as an *L-R* circuit. The current decay curve can be calculated by modelling the instantaneous resistances in the HTS coil, which come from the dynamic resistance, $R_{dyn}$ and joint resistance, $R_j$. For each step in the circuit model, the interval of $I_0 \rightarrow I_1$ is artificially selected by the required accuracy. Then, the decay time, $\Delta t$, is calculated by the $R_{dyn}$ resulting from the FEM model.

To calculate the $R_{dyn}$ value at a certain point of a current decay curve, the intended transport currents in the HTS coil and copper coils are applied using pointwise constraints [23]. In FEM model, $I_{DC}$ in the HTS coil is reached after undergoing a 1 s ramping function and kept constant for 2 s. Then, the external AC field is generated by two copper coils.

This study calculates $R_{dyn}$ within the HTS coil using the average electric field. For 2D axisymmetric coils, the resistive voltage (V) can be calculated by only one-time integration of the entire HTS volume [24]:

$$U_{coil}(t) = \frac{\sum_{n=1}^{N} 2\pi r_n \left( \int_s E_\phi(t) ds \right)}{s} = \frac{\int_{\Omega_{SC}} E_\phi(t) dV}{s} \quad (2)$$

where *s* and $r_n$ are the cross-sectional area and radius of each homogenised turn, $E_\Phi$ is the azimuthal electric field and $\Omega_{SC}$ is the entire superconducting volume. Hence, the dynamic resistance (Ω) can be calculated as:

$$R_{dyn} = f \cdot \int_0^T \frac{U_{coil}(t)}{I_{DC}(t)} dt \quad (3)$$

where *T* is the period and *f* is the frequency of AC field.

## 2.4 Basic feasibility test

In section 2.3, the $I_{DC}$ in the HTS coil, is applied after undergoing a ramping function. However, in the field decay experiment, $I_{DC}$ is supplied by the inductance of the HTS coil and is a decaying function. Therefore, the influence of the differing history of the $I_{DC}$ on the calculation of $R_{dyn}$ must be investigated.

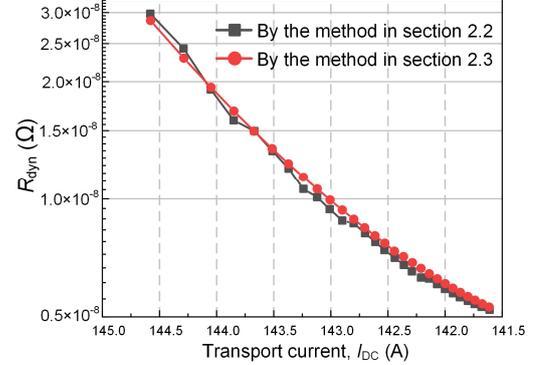

**Figure 4.** The $R_{dyn}$ values calculated by the methods in sections 2.2 and 2.3, as a function of the transport current in HTS coil.

The $R_{dyn}$ values calculated by the method in section 2.3 were compared to which in section 2.2, because the history of the $I_{DC}$ in the latter simulation is identical to the experiment cases. The case study of the single-turn coil in section 2.2 is also performed by the method in section 2.3. For the method in section 2.3, $I_{DC}$ is ramped to the period-average current of each cycle in the current decay curve of figure 3. The charging time and number of AC field cycles are also kept identical to those in figure 3. For the method in section 2.2, $R_{dyn}$ values during current decay can also be obtained using equations (2) and (3). In figure 4, the $R_{dyn}$ values calculated by the two methods are compared, and good agreement can be observed, proving that the simulation in section 2.3 is feasible for estimating $R_{dyn}$ during the current decay.

## 3. Current relaxation

### 3.1 Descending $R_{dyn}$ and current relaxation

The $R_{dyn}$ values of our experimental multi-turn HTS coil are calculated. The detailed coil system is introduced in section 4. $I_{DC}$ in HTS coil and $I_{em}$, which is the RMS current of each copper coil, are applied and kept their values of 55.38 A and 69.3 A, respectively. In figure 5, the coil voltage and $R_{dyn}$ values in the HTS coil are plotted as a function of the cycle number of 50 Hz AC field. However, a problematic phenomenon occurs: the voltage peaks and $R_{dyn}$ values undergo an endless decay as the solution evolves over time. The decay of $R_{dyn}$ did not stop after 100 simulation cycles, and $R_{dyn}$ of the 100th cycle was only 43.8% of the 2nd cycle.

Considering that the experimental field decay process continues for hundreds of seconds (~$10^4$ cycles), we assume that simulation results from larger cycles or steady state will be more related to the experimental conditions. Thus, two questions are posed: (1) Is there a lower limit for the descending $R_{dyn}$? (2) Is there an effective method to determine the stabilised value of $R_{dyn}$ with a reasonable error over limited cycles?

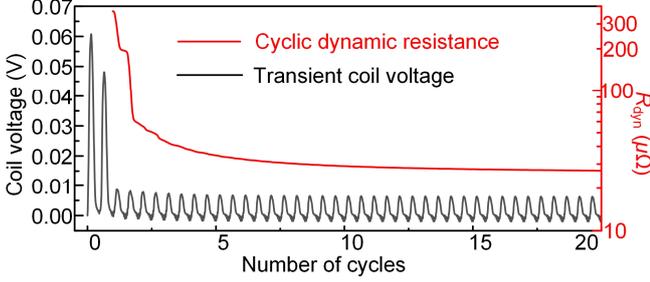

**Figure 5.** Voltage peaks and $R_{dyn}$ values of the multi-turn coil obtained as the simulation cycles increase, when keeping $I_{DC}$=55.38 A and $I_{em}$=69.3 A at 50 Hz.

This phenomenon is concerned with a historical simulation challenge involving the ***E-J*** power law model: current relaxation [17-20]. After applying a pure DC current, the current distribution will always lead to minimal ohmic heat generation. If there is a non-zero ***J***, the power law resistivity will never be zero. Same with normal conductors, DC current will flow from high-resistivity branches to low-resistivity branches. Therefore, current density will penetrate from the edge to the centre of tape and tends to present a homogeneous distribution [17-20], lowering the DC resistance of the entire tape.

In this study, the DC current density under external AC field also tends to penetrate toward the centre of the coil cross-section, causing $R_{dyn}$ values descending with cycles. However, in this case, the current relaxation can be somewhat different from the pure DC current cases. The DC current penetration under AC field is driven by not only the power law resistivity, but also dynamic resistivity.

*3.2 How to deal with current relaxation*

Although the decay of resistance is endless, the impact of current relaxation and descending $R_{dyn}$ on the final results can be very inconsequential by carrying enough number of AC cycles.

The simulation presented in section 3.1 was extended to 100 cycles. In figure 6, the current distribution of the cross-section of HTS coil at the $3/2\pi$ phase of the last cycle is presented. The direction of the relaxation of DC current is illustrated by the black arrows. And the dashed boxes denote the selected innermost/outermost (figure 6(a)), and middle part turns (figure 6(b)), respectively. Their calculated $R_{dyn}$ as functions of simulation cycles are presented in the right two figures, respectively. In figure 6(a), for the innermost /outermost turns, the decay of the $R_{dyn}$ in the first few cycles originates from the superconductors' transitional phase of initial-magnetisation. Particularly, in subsequent cycles, $R_{dyn}$ will stabilise after the current density saturates the cross-section of tape, because the current relaxation has stopped. However, in figure 6(b), for the middle part turns, the $R_{dyn}$ curve still has a negative derivative at the 100$^{th}$ cycle, because the current virgin region (where $J$→0) for further penetration is still large. At this time, the descending of $R_{dyn}$ and variation of current profile is extremely slow, and achieving a completely stable state may require thousands of simulation cycles and unimaginable computational loads.

Thus, the lower limit of the $R_{dyn}$ of the coil can be defined by the stabilised $R_{dyn}$ of the turns shown in figure 6(a). Meanwhile, the descending portion of $R_{dyn}$ of the coil is only contributed by the middle part turns shown in figure 6(b), which determines the error between the descending $R_{dyn}$ and stabilised $R_{dyn}$ of the coil. The magnitude of the error caused by these turns must be investigated.

In figure 6(b), for the turns that $R_{dyn}$ still under descending, a current shielded region ($J$→0) exists in the centre of the tape. Thus, the flux flow cannot cross the central region and exits the conductor from the same side as it enters. Then, the electrical potential caused by flux moving will cancel over a cycle, so the DC electric field would not occur. This indicates that these turns have not reached the threshold states for generating dynamic resistance [4, 15]. In contrast, in figure 6(a), for the turns that $R_{dyn}$ has stabilised, the central region is no longer shielded, indicating that a net flux flow would move across the entire DC region, leading to the development of $R_{dyn}$ [15]. The 100$^{th}$ cycle's $R_{dyn}$ value of the 48 turns in figure 6(b) is only 1.6% of the 8 turns in figure 6(a). So, although the resistance of the turns in figure 6(b) continues to descend, its influence on the resistance of the whole coil is insignificant. The $R_{dyn}$ values of the entire coil calculated by the 100$^{th}$ cycle are very close to the completely stabilised values, thus proving its effectiveness.

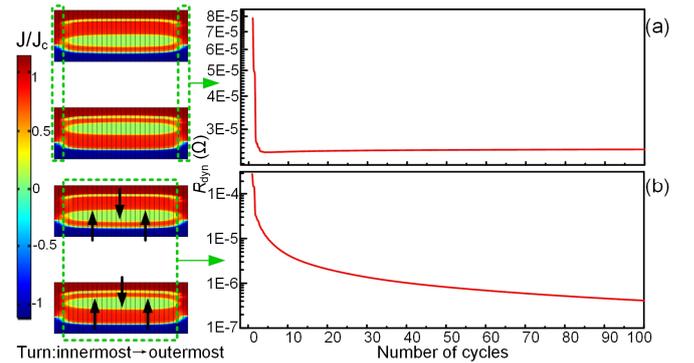

**Figure 6.** (a) Innermost and outermost turns (8 turns); (b) Middle part turns (48 turns). Left figures: current density distribution of the cross-section of the HTS coil at the 100$^{th}$ cycle; Right figures: calculated $R_{dyn}$ as a function of the cycle number, when $I_{DC}$=55.38 A, $I_{em}$=69.3 A at 50 Hz. The black arrows denote the direction of current relaxation.

**4. Experimental verification**

### 4.1 Coil parameters

The experimental HTS coil was wound with GdBCO tape with a width of 6 mm and a thickness of 190 $\mu$m, provided by Shanghai Superconductor Technology Co., Ltd. The thickness of the tape with Kapton insulation was 440 $\mu$m and the total length was about 20 m. Thickness and width of the HTS layer were about 1 $\mu$m, 4.75 mm, respectively. The DP coil, with inner/outer diameters of 10 cm/12.46 cm and a height of 2.1 cm, has 28 turns for each pancake. The coil constant at the centre is about 0.62 mT/A, and the coil $I_c$ is 135 A. The two terminations of the coil are soldered together by a thin tin layer to form a joint with a resistance of $R_j$=12.5 n$\Omega$. The terminals are arranged close to the coil surface to avoid non-axisymmetric geometry that deviates from our model.

### 4.2 Field decay rate measurements

The geometry of the field decay system is shown in figure 7. First, a current of 90 A was excited in the closed-loop coil by a heat-triggered persistent current switch (PCS). Then the HTS coil was operated in the PCM. A Hall sensor was fixed at the centre of the HTS coil to monitor the field and current. Then, the external 50 Hz magnetic field was generated by two identical copper coils. Each copper coil had 317 turns and was powered by a single-phase transformer.

To avoid the influence of the induced AC current circulating in the closed-loop coil, two copper coils were connected anti-parallel to the transformer and symmetrically placed on both sides of the HTS coil. In this way, the axial magnetic fluxes, $B_z$, generated by the two copper coils are in opposite directions and the net $B_z$ variation in the HTS coil was zero. Besides, an epoxy resin sealed XPS box was used to fix the HTS coil and isolates it from the temperature and mechanical disturbances from copper coils.

Figure 8 shows the measured field decay results under ten different AC fields, which are represented by the RMS currents of each copper coils, $I_{em}$. The calculated radial field $B_r$ that is perpendicular to the tape surface generated by unit, $I_{em}$, was 0.98 mT/A at the innermost turn and 1.04 mT/A at the outermost turn. It is clear that the field decay rates increase with the rise of the amplitude of AC fields.

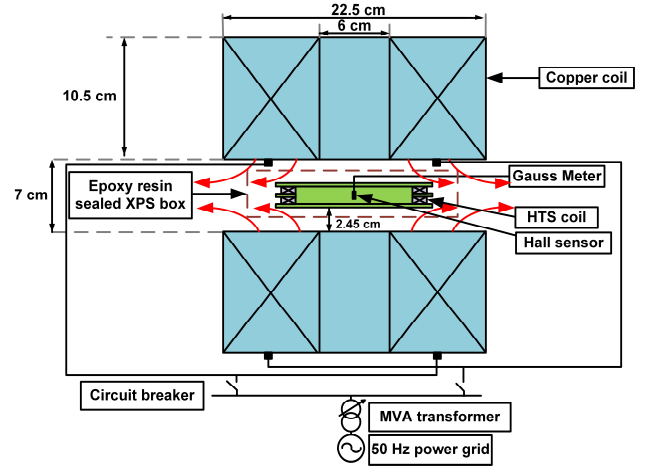

**Figure 7.** Field decay experiment system configuration.

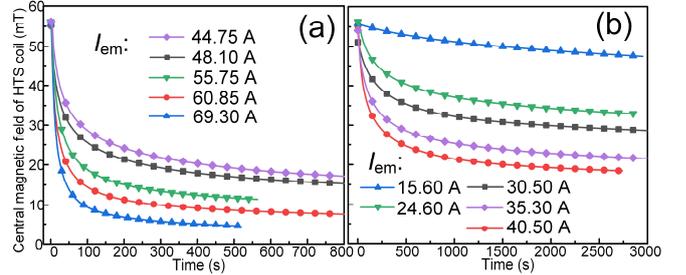

**Figure 8.** Field decay curves under different 50 Hz external fields.

With the calculated coil inductance $L$=0.51 mH, the instantaneous $R_{dyn}$ during decays can be calculated by assuming the HTS coil to be an $L$-$R$ circuit [8]:

$$R_{dyn}(t) = -\frac{dB_z(t)}{dt} L / B_z(t) - R_j \qquad (4)$$

### 4.3 Comparison of experimental and simulated results

The experimental $R_{dyn}$ is calculated by (4). Each simulated $R_{dyn}$ will undergo 100 cycles' current relaxation. Five transport currents, $I_{DC}$, were randomly selected from the experimental current decay curves, and then, examined in the simulation. Figures 9(a)-(e) display $R_{dyn}$ as a function of the external AC field (represented by $I_{em}$), at different $I_{DC}$ in HTS coil, including both the simulated and experimental results. Good agreement is found between the measured and simulated results. It is evident that there exist threshold magnetic fields, $B_{th}$, above which $R_{dyn}$ is generated. The external AC field would have little effect on accelerating the field decay rate if their amplitudes are below $B_{th}$.

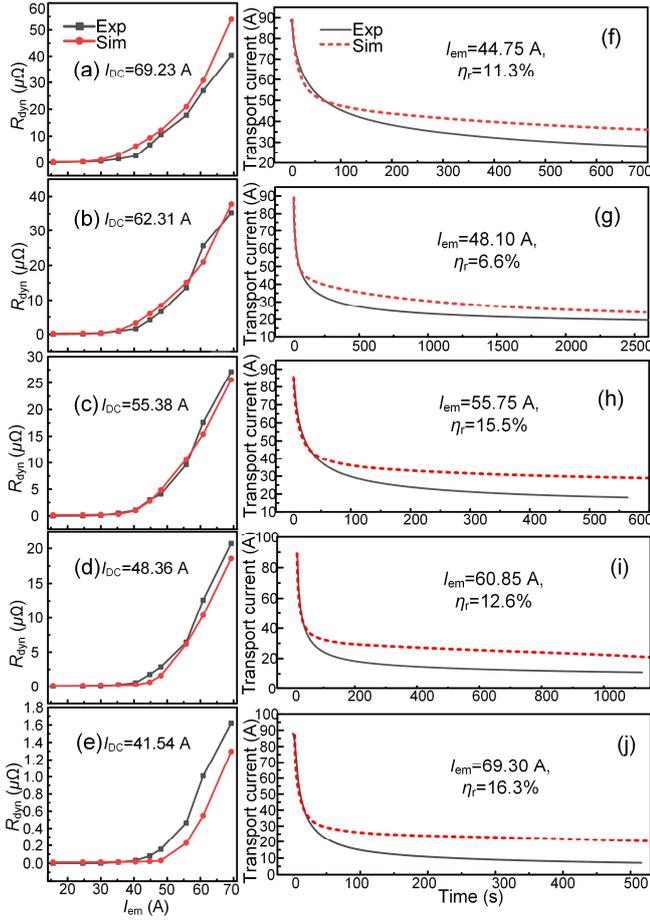

**Figure 9.** (a-e) Measured and simulated $R_{dyn}$ values of the HTS coil at different $I_{DC}$, as a function of the 50 Hz external field. (f-j) Measured and simulated results of current decay curves of the HTS coil under different 50 Hz fields. Amplitude of the external field is represented by the RMS current of each copper coil, $I_{em}$.

The current decay curves are calculated using the technique presented in section 2.3. The initial current of $I_{DC}$ is 89 A, and the interval of $I_0 \rightarrow I_1$ for each step is set as 1 A. To reduce the computational load, $R_{dyn}$ is calculated only at $I_{DC}$ from 89 A to the final value of each decay with an interval of 5 A, and $R_{dyn}$ at other $I_{DC}$ values is obtained by linear interpolation. The experimental conditions in figure 8(a) are examined in simulation. The simulated current decay curves and their comparison with the experimental results are shown in figures 9(f)-(j). Good agreement is found in relatively high-current ranges. However, there is less consistency in the low-current regions of the current decay curves. This difference is likely due to the interactions between the induced AC current and original transport current in the HTS coil. In experiment, the two copper coils were difficult to be manufactured and placed in perfect symmetry to avoid induced current. An induced current caused by $B_z$ also leads to a subtle additional field decay [5, 8], which would become the main contributor for the field decay in the low-current region, where $R_{dyn}$ is negligible. Nevertheless, the maximum relative error of the field decay rates of simulation compared to experiment, $\eta_r$, is within 16.3 %.

## 5. Discussion

### 5.1 $R_{dyn}$ distribution among turns

Figure 10 shows the calculated 100th cycle's $R_{dyn}$ distribution among the turns of the upper pancake of our experimental HTS coil, when $I_{DC}$=69.23 A under different AC fields. Interestingly, most proportion of $R_{dyn}$ is distributed in the innermost and outermost turns, while $R_{dyn}$ of the middle part turns are less than $2 \times 10^{-8}$ Ω. With a portion of the DC and AC fields shielded by the innermost and outermost turns, the middle part turns have higher $I_c$ and are able to shield their DC transport current from external field variations; therefore, little dynamic resistance generated.

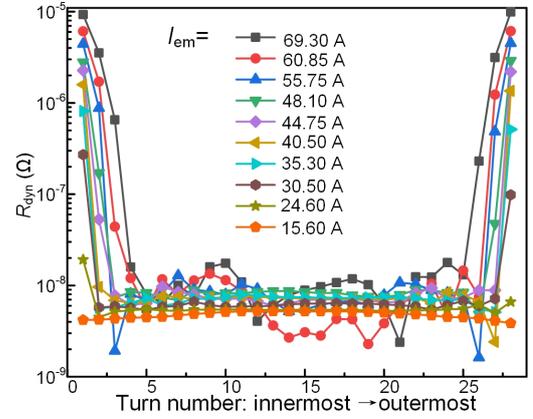

**Figure 10.** $R_{dyn}$ distributions among turns of the HTS coil at the 100th cycle with $I_{DC}$=69.23 A under different amplitudes of a 50 Hz field.

This simulation suggests that the field decay rate is mainly caused by the $R_{dyn}$ of the innermost and outermost turns. Accordingly, higher $I_c$ tapes or particular magnetic shield should be equipped at these locations to reduce the field decay rates.

### 5.2 Magnetic shielding design method

The effectiveness of this magnetic shielding design was tested in simulation. In figure 11(a), two unclosed turns of superconductor tapes are added in the innermost and outermost parts of the 2×28 coil, respectively. In figure 11(b), the significant reduction of field decay rate caused by this shielding design is about 25%. With this model, more efficient shielding (either ferro-magnetic material [7] or HTS tapes) can be designed by varying the number of shielding turns and their space distributions. Even more, HTS tapes with larger width may have better efficiency to shield external field variations. The effectiveness of the shielding technique in the outermost layer of the HTS coil has been tested experimentally [8].

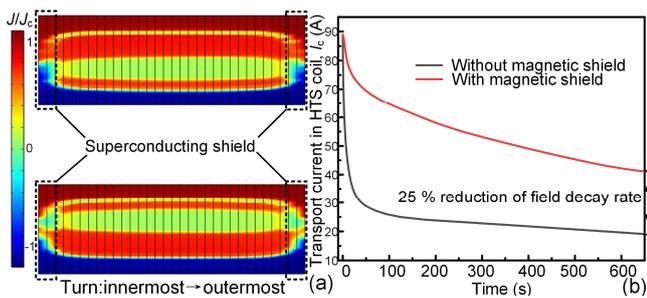

**Figure 11.** (a) Normalised current distribution of the HTS coil with superconducting shield at the $3/2\pi$ phase of the $100^{th}$ cycle, when $I_{DC}$= 54 A and $I_{em}$=69.30 A. (b) Current decay comparison between with and without the superconducting shield, when $I_{em}$=69.30 A.

## 6. Conclusion

Field decay performances in various AC fields are of great importance for the operation of closed-loop HTS magnets. In this work, ***H***-formulation models to calculate the field decay curve affected by dynamic resistance is demonstrated. Feasibility of this model has been tested by comparing with a field cooling and decay model under AC field. The impact of current relaxation on the $R_{dyn}$ results is eliminated by simulating with enough number of AC cycles, at which time $R_{dyn}$ of the turns still under relaxation is negligible, because the threshold for generating $R_{dyn}$ was not reached. The simulated current decay curves and $R_{dyn}$ values show reasonable consistence with the experimental results, although the subtle additional current decay caused by the induced current was not completely eliminated in the experiment. This model suggests that the field decay in AC fields is mainly caused by the dynamic resistance of innermost and outermost turns, so higher $I_c$ tapes or particular magnetic shield structure should be equipped at these locations. This shielding design was examined in simulation and proved to reduce the field decay rate significantly. With this model, the behaviour of closed-loop HTS in AC fields is predicable, and more efficient shielding can be designed by varying the width and number of shielding turns, as well as their spatial distributions.

## Acknowledgements

This work is supported by The National Natural Science Foundation of China (NSFC) under Project 51977130. The authors would like to thank Mr. Xin Yu from Shanghai Superconductor Technology Co., Ltd. for helping in preparing coil samples.